\documentclass[preprint,showpacs,preprintnumbers,amsmath,amssymb]{revtex4}

\usepackage {amssymb}
\usepackage {amsmath}
\usepackage {graphicx}
\usepackage{bm}
\begin{document}
\title{Radiation-induced solitary waves in hot plasmas}

\author{F.V.Prigara}

\affiliation{Institute of Microelectronics and Informatics,
Russian Academy of Sciences,\\ 21 Universitetskaya, 150007
Yaroslavl, Russia} \email{fprigara@imras.yar.ru}

\date{\today}

\begin{abstract}

It is argued that, in a hot plasma interacting with thermal radiation, there
exist the radiation-induced solitary waves different from the types of
nonlinear waves in plasmas known so far. There are extensive observational
evidence for the propagation of these specific density waves in the hot
plasmas of various astrophysical objects. Radiation produces also
non-adiabatic density perturbations analogues to the second sound in the
liquid helium.
\end{abstract}


\maketitle

In hot plasmas, the interaction between radiation and matter plays a very
important role. In fact, a hot plasma is a four-component medium which
includes the electron and ion fluids, magnetic field, and radiation. Each of
the first three components produces its own branch of linear and nonlinear
waves in plasmas [1,2]. Plasma waves and Langmuir envelope solitons
correspond to the electron fluid. Ion sound and ion-sound shocks correspond
to the ion fluid. Magnetic field is responsible for magnetosonic and Alfven
waves, and also for the magnetosonic shocks and solitary waves [3].

Quite similarly, the radiation component of a hot plasma produces its own
branches of linear and nonlinear waves. Linear radiation-induced waves are
non-adiabatic density perturbations, or entropy density waves, similar to
the second sound in a superfluid [4].Here the density of entropy can be
interpreted as the number of photons per one nuclon so as in the Friedmann's
cosmological models with hot matter [5,6].

The nonlinear branch includes radiation-induced density waves in hot plasmas
which are considered in more detail below.

To introduce the radiation-induced density waves in hot plasmas, consider at
first the energy distribution of atoms (ions) in the field of thermal
radiation. We restrict themselves to the case of the sufficiently rarefied
plasma when the line broadening due to the gas pressure is negligible.

Recently, the energy distribution of atoms in the field of thermal blackbody
radiation was obtained [7] in the form

\begin{equation}
\label{eq1}
N/N_{0} = \sigma _{a} \omega ^{2}/\left( {2\pi c^{2}} \right)\left(
{exp\left( {\hbar \omega /kT} \right) - 1} \right),
\end{equation}

\noindent
where $N_{0} $ is the population of the ground state $E_{0} $, \textit{N} is
the population of the energy level $E = E_{0} + \hbar \omega $, $\sigma _{a}
$ is the absorption cross-section, $\hbar $ is the Planck constant, and
\textit{T} is the radiation temperature.

This distribution is valid in the range $\hbar \omega /kT \geqslant 1$,
since in the limit $\hbar \omega /kT \to 0$ the line width is going to
infinity, that indicates the violation of the one-particle approximation
used in [7].

The function (\ref{eq1}) has a maximum at $\hbar \omega _{m} = 1.6kT$. When the
temperature exceeds the critical value of $T_{0} = 3 \times 10^{7^{}}K$ (the
inversion temperature), the population of the energy level \textit{E}
exceeds the population of the ground state $E_{0} $. Since the function (\ref{eq1})
is increasing in the range $\omega < \omega _{m} $ , the inversion of the
energy level population is produced also in some vicinity of $\omega _{m} $
(below $\omega _{m} $). This suggests the maser amplification of thermal
radio emission in continuum by a hot plasma with the temperature exceeding
the critical value $T_{0} $. Maser amplification in compact radio sources
was assumed earlier in [8] based on the high brightness temperatures of
active galactic nuclei. Since a hot plasma in an accretion disk is
concentrated nearby the central energy source, maser amplification is
characteristic for compact radio sources.

It is clear that, when the temperature of a plasma is below $T_{0} $, the
radio flux is very small, and when the temperature exceeds $T_{0} $, radio
emission is on. This an on-off cycle is detected in the radio pulsar PSR
B1259-63 [9]. Similar is an on-off cycle in X-ray pulsars, e.g., the 35-day
cycle in Her X-1. It implies that X-ray emission from X-ray pulsars is
produced by the laser amplification in continuum which is quite analogues to
maser amplification at radio wavelengths.

It was shown recently [7] that thermal emission has a stimulated character.
According to this conception thermal emission from non-uniform gas is
produced by an ensemble of individual emitters. Each of these emitters is an
elementary resonator the size of which has an order of magnitude of mean
free path \textit{l} of photons

\begin{equation}
\label{eq2}
l = \frac{{1}}{{n\sigma} }
\end{equation}

\noindent
where \textit{n} is the number density of particles and $\sigma $ is the
absorption cross-section.

The emission of each elementary resonator is coherent, with the wavelength

\begin{equation}
\label{eq3}
\lambda = l,
\end{equation}

\noindent
and thermal emission of gaseous layer is incoherent sum of radiation
produced by individual emitters.

An elementary resonator emits in the direction opposite to the direction of
the density gradient. The wall of the resonator corresponding to the lower
density is half-transparent due to the decrease of absorption with the
decreasing gas density.

The condition (\ref{eq3}) implies that the radiation with the wavelength $\lambda $
is produced by the gaseous layer with the definite number density of
particles \textit{n} .

The condition (\ref{eq3}) is consistent with the experimental results by Looney and
Brown on the excitation of plasma waves by electron beam (see [1,10]). The
wavelength of standing wave with the Langmuir frequency of oscillations
depends on the density as predicted by equation (\ref{eq2}). The discrete spectrum
of oscillations is produced by the non-uniformity of plasma and the
readjustment of the wavelength to the length of resonator. From the results
of experiment by Looney and Brown the absorption cross-section for plasma
can be evaluated.

The product of the wavelength by density is weakly increasing with the
increase of density. This may imply the weak dependence of the size of
elementary resonator in terms of the wavelength upon the density or,
equivalently, wavelength.

If the temperature of plasma exceeds the inversion temperature, $T_{0} $,
then the inversion of energy level population is created. In this case
thermal radiation from an elementary resonator is amplified by the laser
mechanism and emitted in the direction of decreasing gas density, as
earlier.

The time of life for an elementary resonator has an order of magnitude

\begin{equation}
\label{eq4}
\tau \cong l/v_{iT} \cong l/c_{s} ,
\end{equation}

\noindent
where $v_{iT} $ is the ion thermal velocity, and $c_{s} $ is the speed of
sound. Such is the time

\noindent
duration of the light pulse produced by an elementary resonator.

The absorption of this radiation pulse by the nearby gaseous layer leads to
the heating of a plasma and its density perturbations. The radiation pulse
is then re-emitted by the new virtual elementary resonator in a plasma. In
such a manner, the radiation-induced density wave propagates through the hot
plasma. The width of the density wave has an order of magnitude

\begin{equation}
\label{eq5}
\Delta \cong c\tau \cong cl/c_{s} ,
\end{equation}

\noindent
where \textit{c} is the speed of light.

Various astrophysical objects, such as active galactic nuclei, X-ray
binaries, young pulsars, have hot accretion disks (e.g., [11]). Since the
density gradient in an accretion disk is directed along the radius, the
radiation-induced density wave in a hot plasma of an accretion disk normally
propagates as the radial density wave, though the transverse density wave is
also possible and seems to be observed in some pulsars.

The inversion of energy level population associated with the radial density
wave produces a moving pulse of coherent radiation with the changing
wavelength determined by equations (\ref{eq2}) and (\ref{eq3}). Such are the radio pulses
from pulsars [12].

In fact, the inversion of energy level population corresponding to the radio
band (in some vicinity of $\omega _{m} $ below it - see Sec. 2) exists also
at the temperatures below the inversion temperature, $T_{0} $. This
conclusion is confirmed by the temperature profile in active galactic nuclei
[13] which is virial, i.e.

\begin{equation}
\label{eq6}
T \propto r^{ - 1},
\end{equation}

\noindent
where \textit{r} is the distance from the central energy source. Although
the temperature is decreasing with the increase of the radius, maser
amplification of radio emission producing the high brightness temperatures
is observed in the wide range of wavelengths corresponding, in accordance
with equation (\ref{eq3}), to the wide range of the radius scales.

However, there are no observed pulsed emission from the active galactic
nuclei, contrary to the pulsars. Since the temperature profile in pulsars is
different from those given by equation (\ref{eq6}) and has a form [12]

\begin{equation}
\label{eq7}
dT/dr \geqslant 0,
\end{equation}

\noindent
it is clear, that a hot plasma is required for triggering of radio and X-ray
pulses. Another feature, by which pulsars differ from active galactic
nuclei, is the high ratio of the gas pressure to magnetic pressure (beta) in
pulsars' plasmas [12].

It suggests that the strong magnetic field is not obligatory for the
propagation of the radiation-induced density waves in hot plasmas. However,
if the strong magnetic field is present, it determines the direction in
which the density wave is propagating. Such seems to be the case of solar
microwave bursts. The radiation-induced density wave propagates along a loop
causing the strong correlation in the modulation of radio and X-ray emission
over a large distance ($10^{10}cm$) in the solar corona [14].

Some X-ray binaries, e.g. Sco X-1, show weak emission lines with the
variable intensities and radial velocities. These features are
characteristic for non-saturated lasers. The wavelength of generated mode is
determined by the size of an elementary resonator which is depending on the
density. The radial density wave travelling along the radius changes the
density which results in the variations of the wavelength. Convection or
advection in the gaseous disk produces a two-temperature plasma in which it
is possible to create the inversion of the energy level population. This
gives rise to the weak lasers similar to the weak molecular masers.

The radial density wave produces also the delay of flares at low frequencies
and high frequency quasi-periodic oscillations observed in X-ray binaries.

To summarize, in a hot plasma interacting with thermal radiation, the
radiation-induced solitary waves can propagate in the direction of the
decreasing plasma density. In the presence of the strong magnetic field the
last determines the direction of propagation for the density waves. There is
extensive observational evidence for the density waves in hot plasmas of
accretion disks and the solar corona. The question of detecting the
radiation-induced solitary waves in the hot laboratory plasmas, e.g.
produced by the intense laser beam interacting with a solid surface, remains
open. An effect indicative of the radiation-induced density wave in a hot
plasma is a frequency-shifting pulse of coherent radiation.

\begin{center}
----------------------------------------------------------------------------------------
\end{center}

[1] F.F.Chen, \textit{Introduction to Plasma Physics and Controlled Fusion,
Vol. 1: Plasma Physics} (Plenum Press, New York, 1984).

[2] B.B.Kadomtsev, \textit{Collective Phenomena in Plasmas} (Nauka, Moscow,
1988).

[3] P.K.Shukla, B.Eliasson, M.Marklund, \textit{Nonlinear model
for magnetosonic shocklets in plasmas}, Phys. Plasmas (submitted),
E-print archives, physics/0402035.

[4] L.D.Landau and E.M.Lifshitz, \textit{Fluids Mechanics
}(Butterworth-Heinemann, New York, 1987).

[5] L.B.Okun, \textit{Leptons and Quarks} (Elsevier Science, Amsterdam,
1987).

[6] S.Weinberg, \textit{Gravitation and Cosmology} (Wiley, New York, 1972).

[7] F.V.Prigara, in \textit{Plasmas in the Laboratory and in the
Universe, }September 16-19, 2003, Como, Italy, E-prints archives,
astro-ph/0311532 (2003).

[8] F.V.Prigara, Astron. Nachr.,\textbf{324}, No. S1, 425 (2003).

[9] G.J.Qiao, X.Q.Xue, R.X.Xu, H.G.Wang, and B.W.Xiao, Astron.
Astrophys., \textbf{407}, L25 (2003).

[10] B.V.Alexeev, Usp. Fiz. Nauk,\textbf{173}, 145 (2003),
Physics-Uspekhi, \textbf{46} (2003).

[11] F.Yuan, and A.Zdziarski, \textit{Luminous hot accretion
flows: the origin of X-ray emission of Seyfert galaxies and black
hole binaries, }Month. Not. R. Astron. Soc. (submitted), E-print
archives, astro-ph/0401058 (2004).

[12] F.V.Prigara, in \textit{Gravitation, Cosmology and
Relativistic Astrophysics,} June 23-27, 2003, Kharkov, Ukraine,
E-print archives, astro-ph/0307288 (2003).

[13] N.M.Nagar, A.S.Wilson, and H.Falcke, Astrophys. J.,
\textbf{559}, L87 (2001).

[14] V.V.Grechnev, S.M.White, and M.R.Kundu, Astrophys. J.,
\textbf{588}, 1163 (2003).

\end{document}